\documentclass[prc,showpacs,twocolumn]{revtex4}
\usepackage{graphicx}
\usepackage{dcolumn}
\usepackage{bm}

\begin{document}

\title{Microscopic description of fission in  neutron-rich
Radium isotopes with the Gogny energy density functional}

\author{R. Rodr\'{\i}guez-Guzm\'an$^{1}$ and L. M. Robledo$^{2}$}

\affiliation{
$^{1}$ 
Physics Department, Kuwait University, Kuwait 13060, Kuwait.
\\
$^{2}$ 
Departamento  de F\'{\i}sica Te\'orica, 
Universidad Aut\'onoma de Madrid, 28049-Madrid, Spain
}

\date{\today}

\begin{abstract}

Mean field calculations, based on the D1S, D1N and D1M parametrizations 
of the Gogny energy density functional, have been carried out to obtain 
the potential energy surfaces relevant to fission in several Ra 
isotopes  with the neutron number 144 $\le$ N $\le$ 176. Inner and 
outer  barrier heights as well as first and second isomer excitation 
energies are given. The existence of a well developed third minimum 
along the fission paths of Ra nuclei, is analyzed in terms of the 
energetics of the "fragments" defining such elongated configuration. 
The masses and charges of the fission fragments are studied as 
functions of the neutron number in the parent Ra isotope. The 
comparison between  fission and $\alpha$-decay half-lives, reveals that 
the former  becomes faster for increasing neutron numbers. Though there 
exists a strong variance of the results with respect to the parameters 
used in the computation of the spontaneous fission rate, a change in 
tendency is observed at N=164 with a steady increase that makes heavier 
neutron-rich Ra isotopes stable against fission, diminishing the 
importance of fission recycling in the r-process.
 
\end{abstract}


\maketitle{}

%
%
%

\section{Introduction.}

Fission is a very challenging kind of collective motion whose 
theoretical description can be addressed in terms of the  evolution 
of a given nuclear system from its ground state to scission. To 
simplify this view such evolution is described in terms of several 
intrinsic shapes, labeled by the corresponding deformation parameters 
\cite{Specht,Bjor,Krappe,Baran-Kowal-others-review2015}. The precise 
description of both the fission energy landscape and the associated  
shell effects remains as a major challenge in nuclear structure 
physics, with a potential impact on several basic research and 
technological areas. 

A detailed knowledge of the fission mechanism is required, for example, 
to better understand the very limits of the nuclear stability. As one 
goes up in atomic number Z, the stability of the nucleus against fission 
tends to decrease, due to the increasing Coulomb repulsion, and  the 
quantum  mechanical  shell effects are the only mechanism to increase the 
chances of survival of a given element 
\cite{Cwiok-island,Bender-island,Moeller-island,Warda-Egido-2012}. 

Fission properties of neutron-rich heavy nuclei are also relevant in 
the nucleosynthesis and abundances of elements with mass number greater 
than 120 due to competing fission recycling in the r-process in 
scenarios involving long neutron exposures \cite{Pan05,Mar07,Pan08}. 
Although induced fission is the process to understand, a preliminary 
description of the potential energy surfaces and spontaneous fission 
lifetimes of the involved nuclei can seed some light into the problem. 
Also, systematic studies of the fission paths and related properties, 
based on complementary theoretical approaches, are very useful to 
deepen our knowledge of  the different decay channels (fission, 
$\alpha$-decay, $\dots$)   and their competition in heavy and 
super-heavy nuclei (see, for example, 
\cite{Viola-Seaborg,TDong2005,Rayner-UPRC-2014,Rayner-UEPJA-2014,Robledo-Giulliani} 
and references therein). On the other hand, observables like the 
half-lives, fragment mass and kinetic energy distributions are 
sensitive to the topography of the fission landscape. In turn, an 
improvement in the computation of those quantities would be very useful 
to determine the upper end of the nucleosynthesis flow 
\cite{Arnould-2007}. 

In addition to the physics already mentioned, nuclear fission 
remains a topic of high interest for reactor physics, the degradation 
of radioactive waste, prompt neutron-capture data from  weapon tests 
as well as in the context of the progress made in recent years  in the 
production of super-heavy elements (see, for example, 
\cite{Specht,Krappe,Wagemans,Sierk-PRC2015,JULIN-SHE} and references 
therein). 

With all this in mind, we have carried out fission calculations for the 
radium isotopes from the stability line to very neutron rich isotopes in 
order to study the evolution of the potential energy surfaces and 
spontaneous fission lifetimes. This work can be considered as a 
complement to our previous fission studies in neutron-rich U and Pu 
isotopes \cite{Rayner-UPRC-2014,Rayner-UEPJA-2014}. As a side product 
of the calculations, several examples of second fission isomers have been 
found.

From the theoretical point of view several models are used to describe 
nuclear fission. Among them, both macroscopic-microscopic 
\cite{Sierk-PRC2015,Moller-1,Moller-2,Kowal-1,Kowal-2,Kowal-3,Kowal-4} 
and  mean-field 
\cite{Delaroche-2006,Dubray,Warda-Egido-Robledo-Pomorski-2002,%
Warda-Egido-2012,UNEDF1,Mcdonell-1,Mcdonell-2,Erler2012,Abusara-2010,%
Abu-2012-bheights,RMF-LU-2012,Robledo-Giulliani} ones are  common 
choices. The approximation employed in this work belongs to the second 
class, i.e., the Hartree-Fock-Bogoliubov (HFB) \cite{rs} framework 
based on the highly predictive parametrizations D1S \cite{gogny-d1s}, 
D1N \cite{gogny-d1n} and D1M  \cite{gogny-d1m} of the Gogny 
\cite{gogny} energy density functional (EDF). In this kind of studies, 
the potential energy surfaces (PES) of the different paths to fission, 
the associated collective masses and zero-point quantum corrections 
provided by the mean-field calculations, have been used to compute the 
spontaneous  fission half-lives t$_\mathrm{SF}$ as well as a first 
approach to the masses and charges of the fission fragments based on 
energetics. Special attention has been paid to the uncertainties in the 
computation of the spontaneous fission half-lives and the impact of 
pairing correlations on  observables. For recent complementary work, 
based on the BCPM-EDF \cite{BCMP-EDF}, the reader is  referred to  
\cite{Robledo-Giulliani,Action-Rayner}.

\begin{figure}
\includegraphics[width=0.48\textwidth]{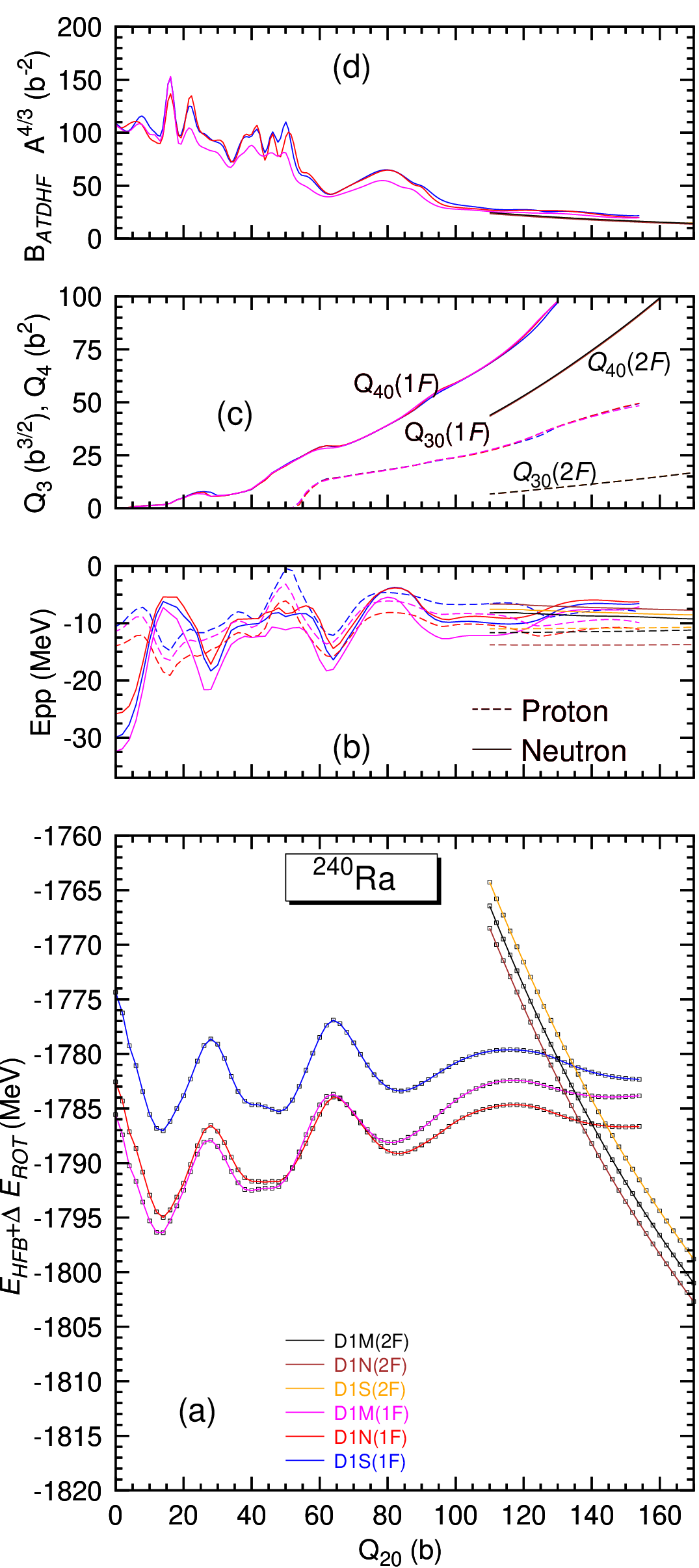}
\caption{ 
(Color online) The HFB plus the zero point rotational energies obtained 
with the D1S, D1N and  D1M  Gogny-EDFs are plotted in panel (a) as functions of the quadrupole 
moment Q$_{20}$ for the nucleus $^{240}$Ra. For each 
EDF, both the one (1F) and two-fragment (2F) solutions 
are included  in the plot. The pairing interaction energies are depicted 
in panel (b) for protons (dashed lines) and neutrons (full lines). 
The octupole and hexadecapole moments 
corresponding to the 1F and 2F solutions are given in panel (c). The collective masses
obtained within the ATDHFB approximation are plotted in panel (d). For more
details, see the main text.
}
\label{FissionBarriersD1SD1ND1M240Ra} 
\end{figure}

In this paper, we focus on the fission properties of even-even Ra 
isotopes with neutron numbers  144 $\le$ N $\le$ 176 in order to 
examine to which  extent  the main features found for the neutron-rich 
U and Pu nuclei are still preserved down to  Z=88. With the aim to 
evaluate the robustness of our predictions, HFB calculations based on 
the D1S, D1N and D1M parametrizations of the Gogny-EDF have been 
performed. It should be kept in mind, that the D1N \cite{gogny-d1n} and 
D1M \cite{gogny-d1m} EDFs provide a better description of the nuclear 
masses, an aspect of relevance to the evaluation of the competing 
$\alpha$ decay channel. In addition, those EDFs were fitted using 
information on neutron matter and therefore they are expected to 
perform well when extrapolating to neutron-rich systems like the ones 
considered in the present study. On the other hand, the thoroughly 
tested Gogny-D1S \cite{gogny-d1s} EDF, already successfully applied to 
fission studies in heavy and super-heavy nuclei 
\cite{Warda-Egido-2012,Delaroche-2006,Warda-Egido-Robledo-Pomorski-2002}, 
has been taken as a reference in this work. This will allow us to test 
the performance of the D1N and D1M Gogny-EDFs scarcely used in fission 
calculations up to now. We implicitly assume that the fission 
properties are determined by general features of the considered 
Gogny-EDFs and therefore, no fine tuning  has been carried out.

The paper is organized as follows. In Sec. \ref{Theory}, we briefly 
outline our theoretical framework. The results obtained for the fission 
paths, barrier heights, fission isomers, fragments' mass and charge in 
$^{232-264}$Ra as well as the isotopic dependence of the spontaneous 
fission half-lives and the competition with the $\alpha$-decay mode are 
discussed in Sec. \ref{results}. Special attention has been paid, in 
the same section, to second fission isomers (i.e., third minima) along 
the fission paths of the considered nuclei. Such minima have  been 
found for several U and Pu nuclei in our recent  HFB studies 
\cite{Rayner-UPRC-2014,Rayner-UEPJA-2014,Robledo-Giulliani} and have 
attracted considerable attention 
\cite{Kowal-3,Kowal-4,3min-Pask,dimolecular-1,dimolecular-2,dimolecular-3,3min-Moller,3min-Cwiok,3min-Beng,3min-Rutz,Delaroche-2006,3min-Berger,Mcdonell-2,3min-Vret}. 
Finally, Sec. \ref{Coclusions} is devoted to the conclusions and work 
perspectives.

\begin{figure*}
\includegraphics[width=1.0\textwidth]{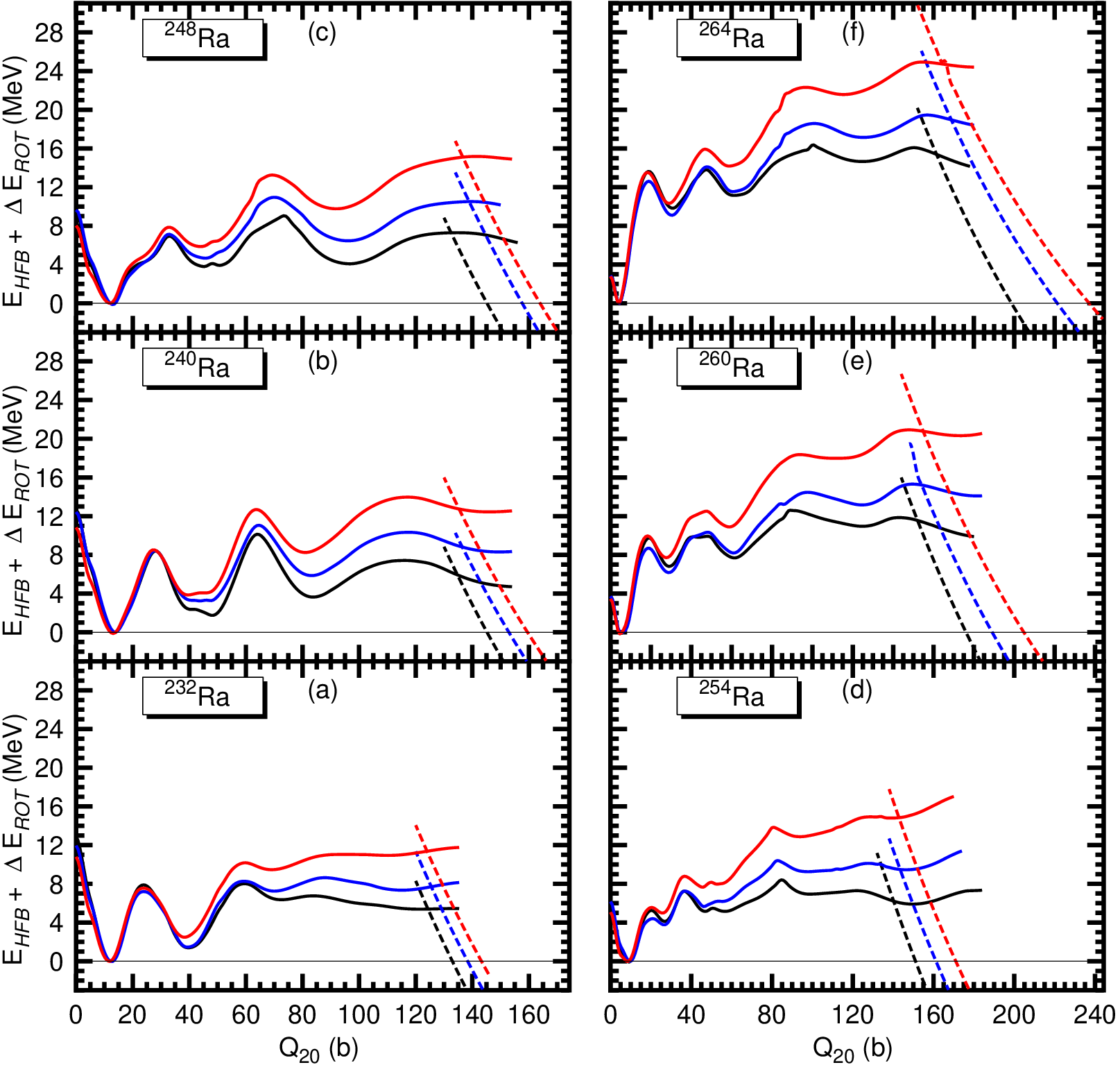} 
\caption{(Color online) The one-fragment 
(full lines)
and two-fragment (dashed lines) HFB plus the zero point rotational energies obtained 
with the 
D1S (black), D1N (blue) and D1M (red) parametrizations of the Gogny-EDF 
are plotted for the nuclei $^{232}$Ra [panel (a)], $^{240}$Ra [panel (b)], $^{248}$Ra [panel (c)],
$^{254}$Ra [panel (d)], $^{260}$Ra [panel (e)] and $^{264}$Ra [panel (f)]
as functions of the quadrupole 
moment $Q_{20}$. All the relative  energies are referred to the absolute minima of the corresponding 
one-fragment curves. For more details, see the main text.
}
\label{FissionBarriers} 
\end{figure*}
 
%
%
%
\section{Theoretical framework}
\label{Theory}

In this section, we briefly outline the theoretical framework used in 
the present study. A detailed account of our methodology can be found 
in   \cite{Rayner-UPRC-2014,Rayner-UEPJA-2014,Robledo-Giulliani}.

The main ingredient is the HFB approximation \cite{rs}, that is used  
with constrains on the axially symmetric quadrupole  $\hat{Q}_{20}$ and 
octupole $\hat{Q}_{30}$ operators 
\cite{Rayner-UPRC-2014,Rayner-UEPJA-2014,Robledo-Giulliani,Action-Rayner,PRCQ2Q3-2012,Robledo-Rayner-JPG-2012} 
to obtain one-fragment (1F) solutions. We are aware of the role played 
by triaxiality for configurations around the top of the inner barrier 
\cite{Rayner-UPRC-2014,Abusara-2010,Delaroche-2006}. However, we have 
kept axial symmetry, as a selfconsistent symmetry, along the whole 
fission path in order to reduce the already substantial computational 
effort. The $\gamma$ degree of freedom has also been neglected in the 
computation of our t$_\mathrm{SF}$ values [see, Eqs.(\ref{TSF}) and 
(\ref{Action}) below] as it has  been shown in previous studies 
\cite{Bender-island,Baran-1981} that its impact is very limited. 

For large values of the quadrupole moment Q$_{20}$, two-fragment (2F) 
solutions have been reached by constraining on the necking operator 
$\hat{Q}_{Neck}(z_{0},C_{0})$. In computing spontaneous fission lifetimes
(see below) the ridge connecting 
the 1F and 2F curves in the multidimensional space of deformations 
($Q_{20}$,$Q_{30}$,$Q_{Neck}$, $\dots$) has been neglected and therefore the 2F curves 
are considered as really intersecting the 1F ones 
\cite{Rayner-UPRC-2014,Rayner-UEPJA-2014,Robledo-Giulliani}. 

\begin{figure}
\includegraphics[width=0.5\textwidth]{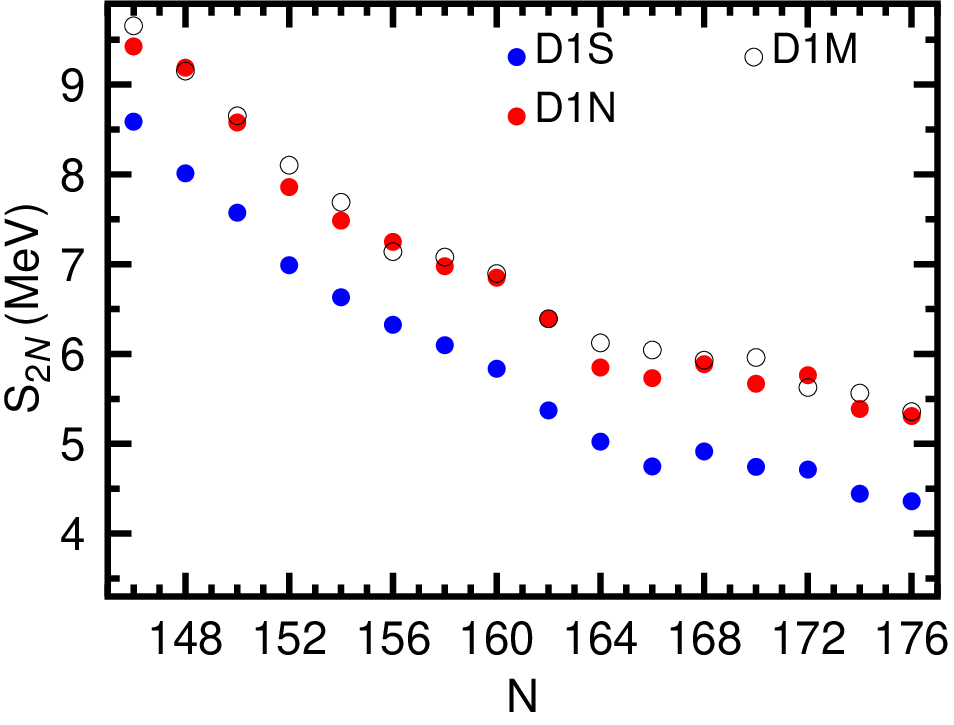}
\caption{ (Color online)  
Two neutron separation energies S$_{2N}$ as  
functions of the neutron number. 
}
\label{S2N-D1S-D1N-D1M} 
\end{figure}

Aside from the constraints already mentioned, as well as the  typical 
HFB ones on both the proton and neutron numbers \cite{rs}, a constraint 
on the operator $\hat{Q}_{10}$ is used  to prevent spurious effects 
associated to the center of mass motion 
\cite{PRCQ2Q3-2012,Robledo-Rayner-JPG-2012}. Note, that the average 
values of higher multipolarity deformations (i.e., $\hat{Q}_{40}$, 
$\hat{Q}_{60}$, $\dots$) are automatically adjusted during the 
selfconsistent minimization of the HFB energies. Furthermore, as a 
result of projecting multidimensional paths into one-dimensional plots 
(see, Figs. \ref{FissionBarriersD1SD1ND1M240Ra}  and 
\ref{FissionBarriers}) kinks and multiple branches may appear in this 
type of calculations 
\cite{Rayner-UPRC-2014,Rayner-UEPJA-2014,Robledo-Giulliani,Dubray-discontinuities}.

The HFB quasiparticle operators are expanded in a deformed axially 
symmetric harmonic oscillator (HO) basis with quantum numbers 
restricted by the condition 
$
2 n_{\perp} + |m| + \frac{1}{q} n_{z} \le \textrm{M}_{z, \textrm{MAX}}	
$
where M$_{z,\textrm{MAX}}$=17 and $q=1.5$. This amounts to
consider states with J$_{z}$ quantum numbers
up to 35/2 and up to 26 quanta in the $z$-direction. For 
a neutron-rich nucleus like $^{280}$Pu this basis provides an 
error (with respect to a larger basis with M$_{z,\textrm{MAX}}$=18) smaller
than 0.8 MeV all over the fission path
\cite{Rayner-UEPJA-2014}. In addition, for each of the considered 
($Q_{20}$,$Q_{30}$,$Q_{Neck}$, $\dots$)-configurations, the 
HO lengths b$_{\perp}$ and b$_{z}$ have been optimized so as
to minimize the total HFB energy. This guarantees a much better 
convergence for  relative  energies
(see, Fig.\ref{FissionBarriers}). 

For the solution of the constrained 
mean-field equations we have employed an 
approximate second order gradient method based on the parametrization 
of a given HFB vacuum with the help of the Thouless theorem 
\cite{PRCQ2Q3-2012,Robledo-Rayner-JPG-2012,Robledo-Bertsch2OGM}. The 
two-body kinetic energy correction has been fully taken into account in 
the Ritz-variational \cite{rs} procedure  while for the Coulomb 
exchange term we have considered the Slater approximation. The 
spin-orbit contribution to the pairing field has been neglected.

We have computed the spontaneous fission half-life   using the 
Wentzel-Kramer-Brillouin (WKB)  formalism \cite{Baran-TSF-1,Baran-TSF-2}
 
\begin{eqnarray} \label{TSF}
t_\mathrm{SF}= 2.86 \times 10^{-21} \times \left(1+ e^{2S} \right)
\end{eqnarray}
where the action $S$ along the (minimal energy one-dimensional projected)
fission path reads
\begin{eqnarray} \label{Action}
S= \int_{a}^{b} dQ_{20} \sqrt{2B(Q_{20})\left(V(Q_{20})-\left(E_\mathrm{GS}+E_{0} \right)  \right)}.
\end{eqnarray}  
In the above expression, $B(Q_{20})$ and $V(Q_{20})$ represent the 
collective mass and potential for the collective variable $Q_{20}$. The 
potential energy is given by the HFB energy of the corresponding 
constrained state corrected by the quantum  zero point vibrational  
$\Delta E_\mathrm{vib}(Q_{20})$ and rotational $\Delta E_\mathrm{ROT}(Q_{20})$ energies. 
Both the inertia $B(Q_{20})$ and $\Delta E_\mathrm{vib}(Q_{20})$ 
have been computed using the  cranking approximation to the Adiabatic 
Time Dependent HFB (ATDHFB) approach 
\cite{crankingAPPROX,Giannoni-1,Giannoni-2,Libert-1999} and the 
Gaussian Overlap Approximation (GOA) to the Generator Coordinate Method 
(GCM) \cite{rs,Rayner-UPRC-2014,Rayner-UEPJA-2014,Robledo-Giulliani} 
while $\Delta E_\mathrm{ROT}(Q_{20})$ has been computed in terms of the Yoccoz 
moment of inertia \cite{RRG23S,ER-Lectures}. The integration limits $a$ 
and $b$ in Eq. (\ref{Action}) are the classical turning points 
\cite{proportional-1} corresponding to the energy $E_\mathrm{GS}+E_{0}$. For 
the free parameter $E_{0}$, we have considered the four values 
$E_{0}$=0.5, 1.0, 1.5 and 2.0 MeV in order to study its impact on lifetimes. 
It should be kept in mind that 
different E$_{0}$ values lead to different  integration limits and, 
therefore,  modify the value of the integral in Eq. (\ref{Action}). This is particularly 
relevant, in the case of  neutron-rich Ra isotopes which display high 
and wide fission barriers. 

Finally, in order to study the competition between the 
spontaneous fission and $\alpha$-decay modes, we have   computed 
the corresponding  $t_{\alpha}$ values  using the Viola-Seaborg formula
\begin{equation} \label{VSeaborg-new}
\log_{10} t_{\alpha} =  \frac{AZ+B}{\sqrt{ {\cal{Q}}_{\alpha}}} + CZ+D
\end{equation} 
with the parameters $A$=1.64062, $B$=-8.54399, $C$=-0.19430 and 
$D$=-33.9054 as given in \cite{TDong2005}. The ${\cal{Q}}_{\alpha}$ 
value  is obtained from the calculated binding energies  for  Ra  and 
Rn nuclei. Obviously, other types of decay (for example, $\beta$-decay) 
may play a role  in the case of heavy neutron-rich nuclei. However, 
their study lies outside the scope of this work.

%
%
%

\section{Discussion of the results}
\label{results}

An illustrative outcome of our calculations is presented in 
Fig.\ref{FissionBarriersD1SD1ND1M240Ra} for the nucleus $^{240}$Ra. A 
similar analysis, as described below, has been carried out for each of 
the studied Ra isotopes. In panel (a), we have plotted the HFB energies 
plus the rotational  corrections E$_{HFB}$+ $\Delta$ E$_\mathrm{ROT}$ as 
functions of the quadrupole moment. The zero point vibrational energies 
are not included in the plot, as they are rather constant as functions 
of Q$_{20}$. However, they are always included in the computation of 
the spontaneous fission half-lives.

\begin{figure}
\includegraphics[width=0.5\textwidth]{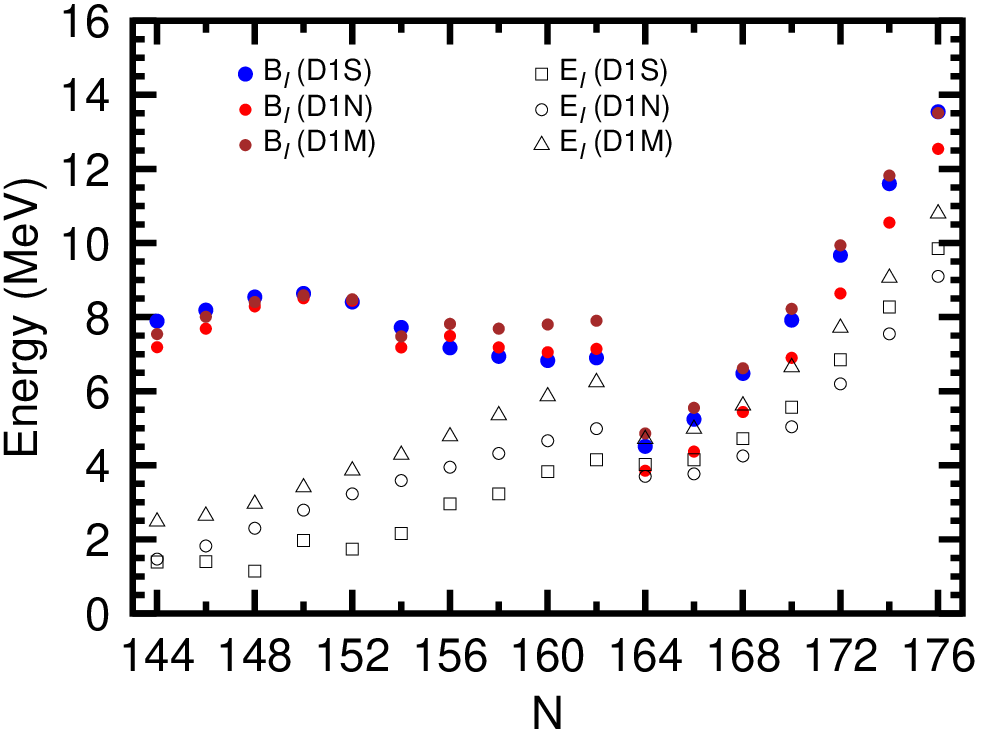}
\caption{ (Color online) Excitation energies $E_{I}$ of the first fission isomers and
inner  barrier heights $B_{I}$  as function 
of the neutron number in $^{232-264}$Ra.
}
\label{Barriers-D1S-D1N-D1M-I} 
\end{figure}

As can be seen, the Gogny-D1S EDF provides a pronounced under-binding as 
compared to the D1N and D1M ones. This reflects a well known deficiency 
of the former in heavier nuclei 
\cite{Rayner-UPRC-2014,Rayner-UEPJA-2014,Hilare-2007} as a result of 
which both the D1N and D1M parametrizations have been specially 
tailored in an effort to build an accurate mass table based on the 
Gogny-EDF \cite{gogny-d1n,gogny-d1m}. A similar behavior has been found 
in our symmetry-projected configuration mixing study of the quadrupole 
collectivity across the N=126 neutron shell closure \cite{RaynerN126} 
as well as in \cite{TomasGlobal}. Nevertheless, it is satisfying to 
observe that the 1F curves provided by all the functionals are rather 
similar with the ground state  at Q$_{20}$= 14 b. 

The first fission isomer appears at Q$_{20}$= 48, 44 and 40 b with the 
D1S, D1N and D1M parametrizations. Their excitation energies are 1.74, 
3.23 and  3.86 MeV, respectively. They are separated from the ground 
state by  inner barriers with heights (without triaxiality) of 8.41, 
8.44 and 8.48 MeV, respectively. Second fission isomers (Q$_{20}$ 
$\approx$ 82 b) are also apparent from panel (a). They lie 3.63, 5.87 
and  8.26 MeV above the ground state while the heights of the second 
barriers (Q$_{20}$ $\approx$ 64 b) are 10.13, 11.03 and  12.67  MeV 
with the D1S, D1N and  D1M parametrizations. Reflection-asymmetric 
second fission isomers, have already  been found in previous studies 
\cite{Rayner-UPRC-2014,Rayner-UEPJA-2014,Robledo-Giulliani,Kowal-3,Kowal-4,3min-Pask,3min-Moller,3min-Cwiok,3min-Beng,3min-Rutz,Delaroche-2006,3min-Berger,Mcdonell-2,3min-Vret}. 
As will be shown later on (see, Fig. \ref{FissionBarriers}) they also 
emerge along the fission paths of several Ra isotopes.

\begin{figure}
\includegraphics[width=0.5\textwidth]{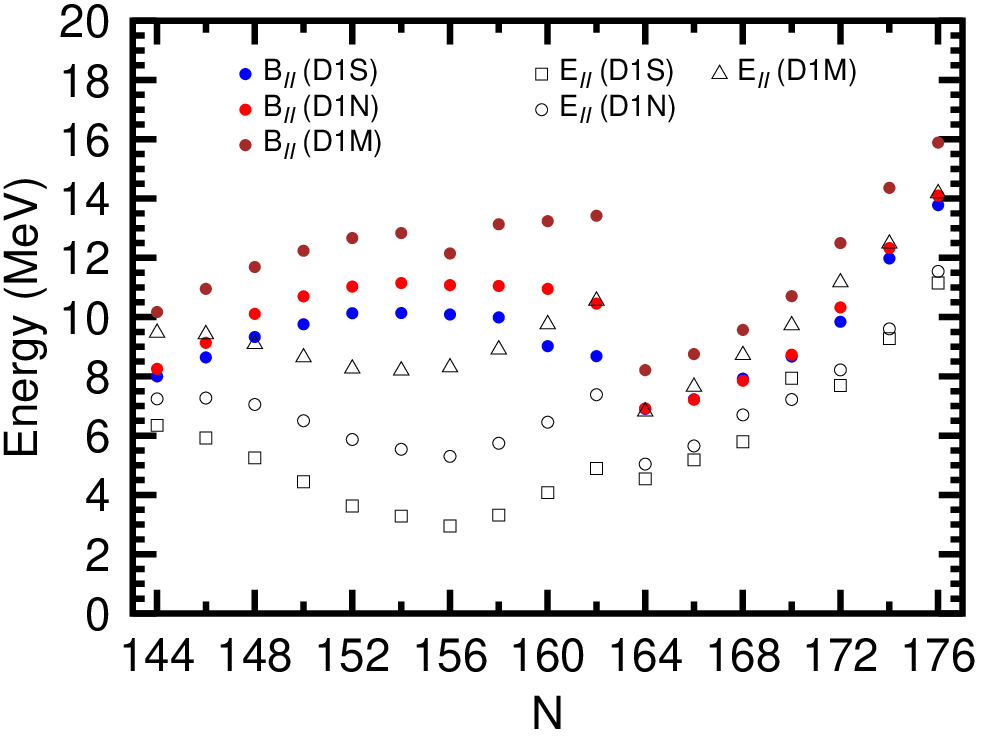}
\caption{ (Color online) Excitation energies $E_{II}$ of the second fission isomers and
second  barrier heights $B_{II}$  as function 
of the neutron number in $^{232-264}$Ra.
}
\label{Barriers-D1S-D1N-D1M-II} 
\end{figure}

The proton (dashed lines) and neutron (full lines) pairing correlation 
energies E$_{pp}$= 1/2 Tr($\Delta \kappa^{*}$) are shown in panel (b). 
Minima are observed for the neutron pairing energies at the spherical 
configuration, the top of the inner and second barriers as well as for 
Q$_{20}$ $\approx$ 100 b. In panel (c), we have plotted the octupole 
Q$_{30}$ and hexadecupole Q$_{40}$ moments corresponding to the 1F 
[i.e., Q$_{30}(1F)$ and Q$_{40}$(1F)] and 2F [i.e., Q$_{30}(2F)$ and 
Q$_{40}$(2F)] paths which are clearly separated in the multidimensional 
(collective) deformation space and rather similar for all the 
parametrizations.

The ATDHFB collective masses are displayed in panel (d). Their behavior 
is well correlated with the one for the pairing energies shown in 
panel (b) and the inverse  dependence of the collective mass with the 
square of the pairing gap \cite{proportional-1,proportional-2}. The GCM 
collective masses (not shown in the figure) display a similar pattern 
though they are smaller than the ATDHFB ones.  For $^{240}$Ra and  
E$_{0}$=1.0 MeV, for example, this leads to pronounced differences of 
10, 12 and 13  orders of magnitude in the t$_\mathrm{SF}$ values predicted in 
the two schemes with the D1S, D1N and D1M parameter sets, respectively. 
These large uncertainties, are one of the main reasons driving our 
choice of both  schemes in the computation of the spontaneous fission 
half-lives. In all the computations of the t$_\mathrm{SF}$ values, the wiggles 
in the collective masses have been softened by means of a three point 
filter \cite{Rayner-UPRC-2014}.

\begin{figure*}
\includegraphics[width=0.31\textwidth]{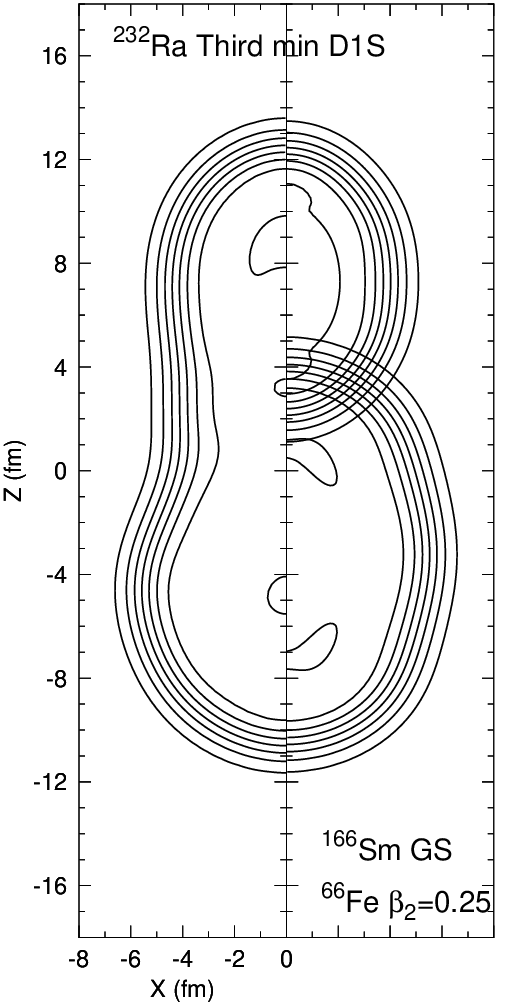}%
\includegraphics[width=0.305\textwidth]{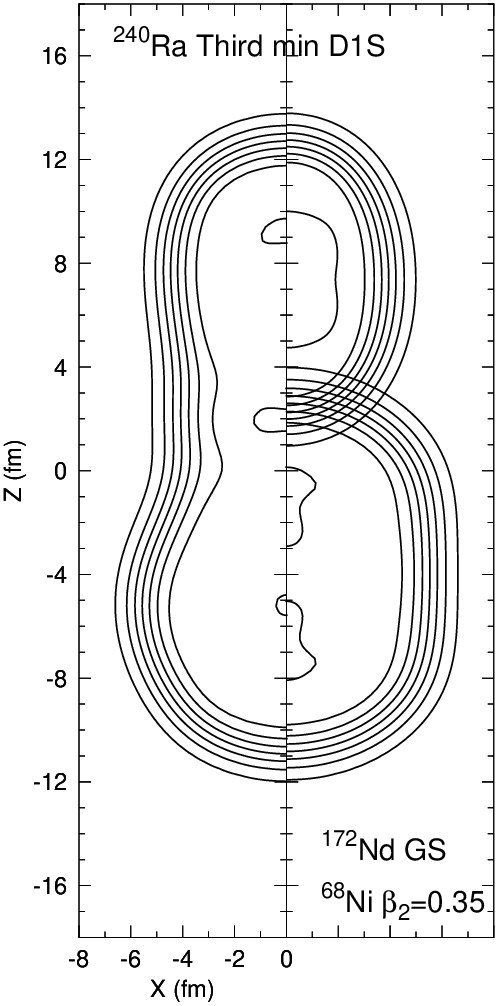}%
\includegraphics[width=0.31\textwidth]{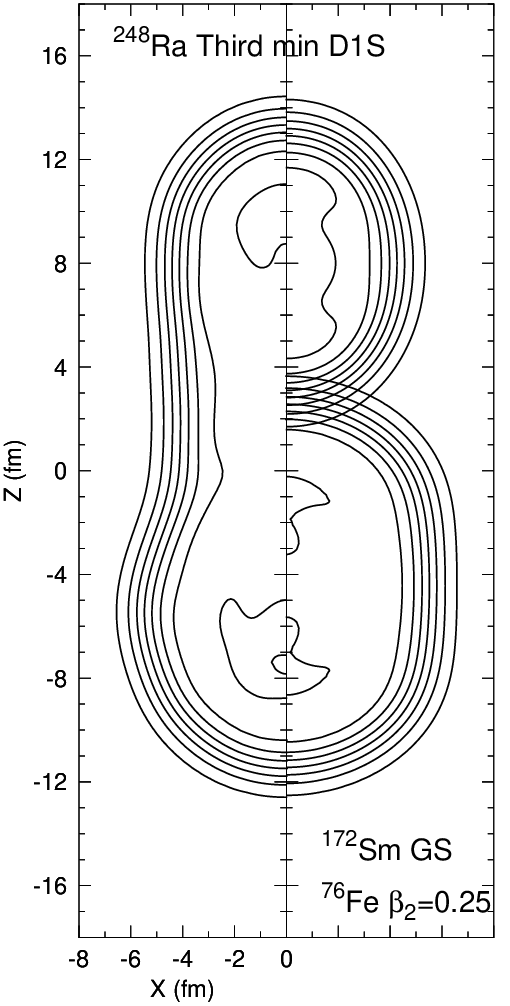}%
\caption{The spatial density  
(left hand side of each panel)
corresponding to the second isomer 
in the nuclei $^{232}$Ra, $^{240}$Ra 
and $^{248}$Ra  is compared to the densities 
(right hand side of each panel)
of two nuclei summing up
the same number of protons and neutrons as the parent nucleus. 
The density is in
units of fm$^{-3}$.
Results are shown for the Gogny-D1S EDF. For more details, see the main text.
 }
\label{CDthird} 
\end{figure*}

Let us also mention that, for large quadrupole moments, the 2F 
solutions in $^{240}$Ra correspond to a spherical $^{130}$Cd  and an 
oblate ($\beta_{2}$=-0.21) and  slightly octupole deformed 
($\beta_{3}$=0.03) $^{110}$Zr  fragment. With the D1S, D1N and  D1M 
Gogny-EDFs, the  oblate $^{110}$Zr fragment minimizes  Coulomb 
repulsion energies of 166.31, 166.34 and  166.55 MeV, respectively. 
Oblate fragments have also been found by fissioning other Ra (see, 
below), U and Pu 
\cite{Rayner-UPRC-2014,Rayner-UEPJA-2014,Robledo-Giulliani} nuclei. 
Here, we would like to stress that  fragment masses are the result of 
the miniminization of the HFB energy at a given distance between the 
fragments and therefore should only be taken as  rough approximations 
to the peaks of the broad fragments' mass distribution 
\cite{Goutte-dynamical-distribution}. The heavier fragment, $^{130}$Cd, 
is a consequence of its neutron number N=82 being a magic number 
\cite{Rayner-UEPJA-2014,Nenoff-2007,Piessens-1993,Ter-1996}. A more 
realistic description of fragments' mass distribution 
\cite{Pu-mass-fragments-exp-2,Pu-mass-fragments-exp-1} has to take into 
account the dynamics of the system around the loosely defined scission 
configuration 
\cite{Rayner-UPRC-2014,Rayner-UEPJA-2014,Chasman-breaking}.

In Fig.~\ref{FissionBarriers}  we have plotted the energies E$_{HFB}$+ 
$\Delta$ E$_\mathrm{ROT}$, as functions of the quadrupole moment 
$Q_{20}$, for $^{232}$Ra, $^{240}$Ra, $^{248}$Ra, $^{254}$Ra, 
$^{260}$Ra and $^{264}$Ra as a representative sample of the considered 
isotopes. Both the 1F (full lines)   and 2F (dashed lines) paths are 
shown in the plots. For each isotope the relative energies are  always 
referred to the absolute minima of the 1F curves in order to 
accommodate all the paths, obtained with the D1S, D1N and D1M 
Gogny-EDFs, in a single plot. Results for $^{240}$Ra are also included 
in the figure for the sake of completeness. 

The ground state deformation decreases with increasing neutron number N 
reaching its minimal value Q$_{20}$= 2b for $^{264}$Ra. Using the 
corresponding ground state energies, we have computed the  two-neutron 
separation energies S$_{2N}$  shown in Fig.~\ref{S2N-D1S-D1N-D1M}. 
Regardless of the considered functional, the S$_{2N}$ values exhibit a 
clear decreasing trend with increasing neutron number. In the case of 
$^{234}$Ra we have obtained S$_{2N}$=8.59, 9.42 and 9.66 MeV with the 
D1S, D1N and D1M parametrizations while for $^{264}$Ra the 
corresponding two-neutron separation energies are S$_{2N}$=4.36, 5.48 
and  5.36 MeV, respectively. The smaller  S$_{2N}$ values obtained with 
the D1S  as compared with the D1N and D1M parametrizations (typically 
around 1 MeV) are due to the well known under-binding of the D1S 
parametrization.

\begin{figure*}
\includegraphics[width=0.985\textwidth]{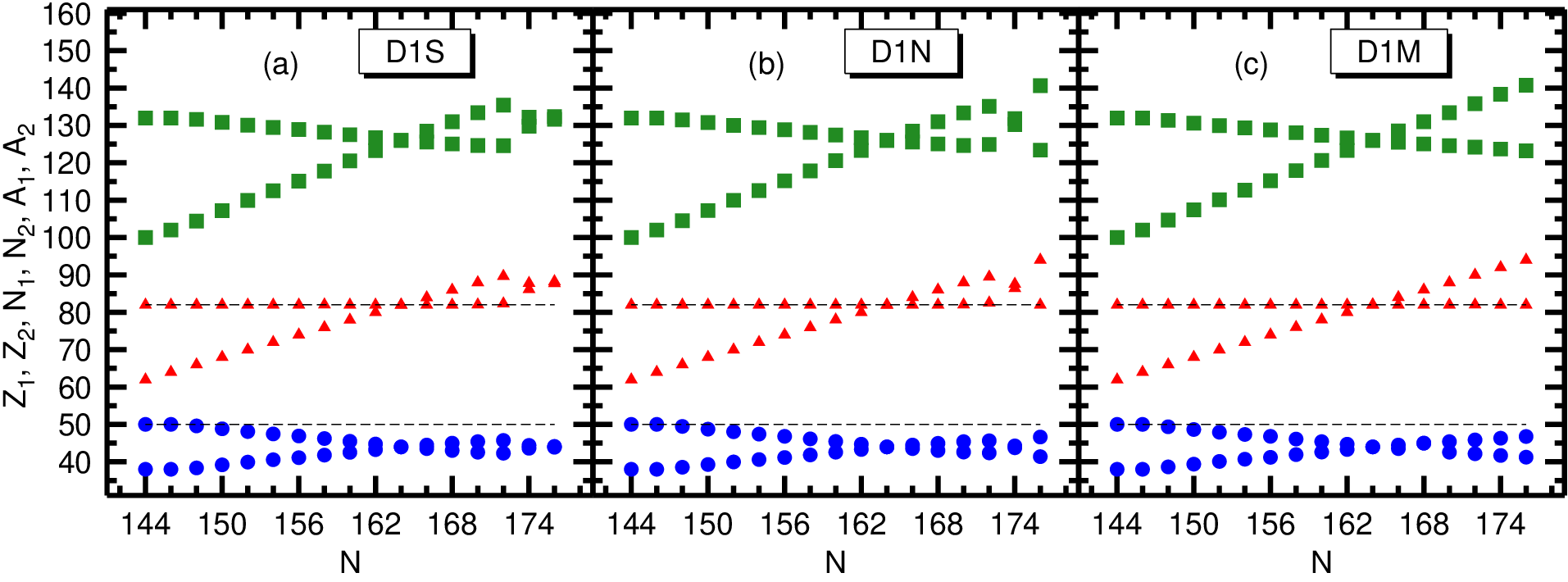} 
\caption{ (Color online) The proton ($Z_{1},Z_{2}$), neutron ($N_{1},N_{2}$) and mass ($A_{1},A_{2}$)
numbers of the two  fragments obtained in our HFB calculations for the isotopes $^{232-264}$Ra 
are shown as functions of the neutron number in the  parent nucleus.
Results have been obtained with the Gogny-D1S [panel (a)], Gogny-D1N [panel (b)] and Gogny-D1M [panel (c)]
EDFs. The magic proton Z=50 and  neutron N=82 numbers
are highlighted with dashed horizontal lines to guide the eye.  
}
\label{mass-D1S-D1N-D1M} 
\end{figure*}

The excitation energies E$_{I}$ of the first fission isomers and the 
inner  barrier heights B$_{I}$ (without triaxiality) are summarized  in 
Fig.~\ref{Barriers-D1S-D1N-D1M-I}. With all the functionals, the 
E$_{I}$ values increase almost linearly as functions of the neutron 
number exception made of $^{252}$Ra (4.02, 3.70 and  4.71 MeV with the 
D1S, D1N and D1M Gogny-EDFs) for which a change in tendency is 
observed. The barrier heights B$_{I}$ reach their minimal values (4.51, 
3.85 and  4.86 MeV with the D1S, D1N and  D1M Gogny-EDFs) for the N=164 
nucleus $^{252}$Ra and increase  for heavier isotopes. Such an increase 
agrees well with the HFB predictions for neutron-rich U and Pu nuclei 
\cite{Rayner-UPRC-2014,Rayner-UEPJA-2014,Robledo-Giulliani} and 
previous   Extended Thomas-Fermi (ETF) results \cite{Mamdouth}. An 
increase is also visible in the recently reported 
macroscopic-microscopic B$_{I}$ values for even-even Ra isotopes with 
166 $\le$ N $\le$ 182 \cite{Sierk-PRC2015}. Larger B$_{I}$ values, 
together with the widening of the 1F curves, leads to larger 
spontaneous fission half-lives as one moves towards more neutron-rich 
systems (see, Fig.~\ref{tsf-D1S-D1N-D1M}).

We are aware of the reduction, by a few MeV, of the inner barrier 
heights B$_{I}$ once triaxiality is taken into account (see, for 
example \cite{Abusara-2010}). In fact, such a reduction has 
already been found in our previous Gogny-D1M HFB calculations for a set 
of U, Pu, Cm and Cf nuclei (see Table I and Fig.~4 of 
 \cite{Rayner-UPRC-2014}). Though we have mainly kept axial symmetry 
along the fission paths of the considered Ra isotopes, we have also 
corroborated in some selected cases (i.e., $^{232}$Ra, $^{236}$Ra and 
$^{244}$Ra) the reduction of the corresponding B$_{I}$  values once the 
$\gamma$ degree of freedom  is included. However, such a reduction 
comes together with an increase of the collective inertia 
\cite{Baran-1981,Bender-1998} that tends to compensate in the final 
value of the action. As a result, the impact of the $\gamma$ degree of 
freedom is very limited and has not been considered in our calculations 
of the spontaneous fission half-lives.

Coming back to  Fig.~\ref{FissionBarriers}, one also observes another 
common  feature  provided by all the the Gogny-EDFs, i.e., second 
fission isomers in the 1F curves of the studied Ra nuclei. For the 
isotopes $^{232-252}$Ra their  quadrupole deformations are within the 
range 68 b $\le$ Q$_{20}$ $\le$ 102 b. They are also apparent in the 1F 
curves of the heavier isotopes though in that case additional shallow  
minima have  been found.  

The second  barrier heights B$_{II}$ and the excitation energies 
E$_{II}$ of the second fission isomers are summarized in 
Fig.~\ref{Barriers-D1S-D1N-D1M-II}. Though all the considered 
functionals provide a similar trend, the largest barrier heights 
B$_{II}$ are obtained with the D1M parametrization. A sudden drop in 
the B$_{II}$ values occurs for the N=164 isotope $^{252}$Ra (6.90, 6.92 
and  8.21 MeV with the D1S, D1N and  D1M Gogny-EDF) followed  by an 
almost linear increase in  heavier isotopes. Note, that for neutron 
numbers N  $\ge$ 164, the predicted D1S and D1N B$_{II}$ values are 
rather close (see, Fig.~\ref{FissionBarriers}). The  largest E$_{II}$ 
energies are also the ones obtained with the Gogny-D1M EDF. Here, one 
observes two minima (one at N=156 and the other at N=164 regardless of 
the EDF employed) and a steady increase in heavier isotopes for which 
the corresponding D1S and D1N  E$_{II}$ values are, once more, rather 
close (see, Fig.~\ref{FissionBarriers}).

The previous results for E$_{I}$, E$_{II}$, B$_{I}$ and B$_{II}$ agree 
well with the ones obtained for U and Pu nuclei  
\cite{Rayner-UPRC-2014,Rayner-UEPJA-2014}. In particular, one sees 
that, regardless of the employed functional, third minima along the 
fission paths  represent a  robust feature within our Gogny-HFB 
framework. They are also visible in the 1F curves of the isotopes 
$^{232-254}$Th for which preliminary calculations have been carried 
out. The same conclusions can be extracted from recent BCPM 
\cite{Robledo-Giulliani}, Skyrme \cite{Mcdonell-2} and  relativistic 
mean-field \cite{3min-Vret} studies. Therefore, the shell effects 
leading to second fission isomers (i.e., third minima) in this region 
of the nuclear chart are systematically present in all the mean-field 
approximations already mentioned. However, further studies are required 
in order to clarify the relation between the mean-field predictions and 
the available experimental data.

\begin{figure*}
\includegraphics[width=0.3\textwidth]{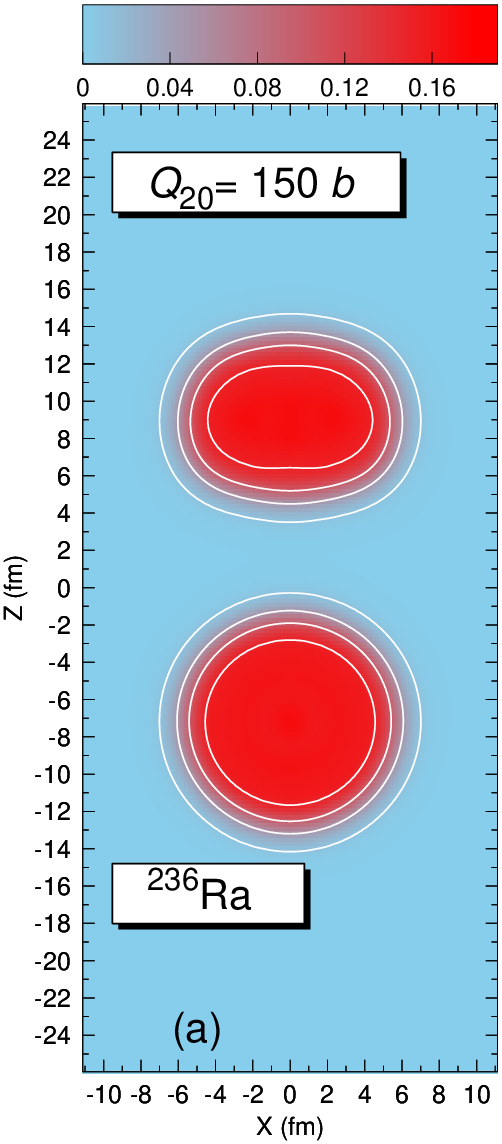}
\includegraphics[width=0.3\textwidth]{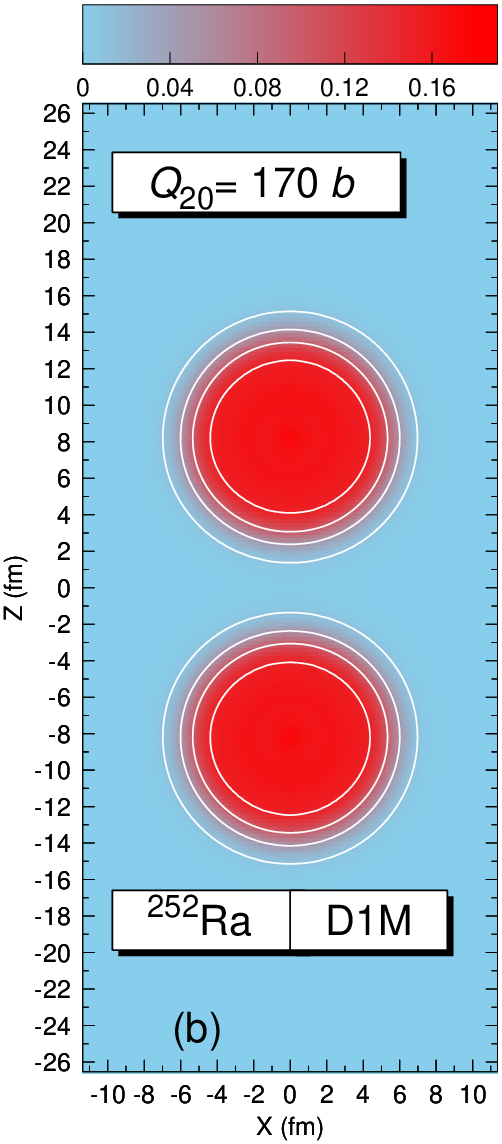}
\includegraphics[width=0.3\textwidth]{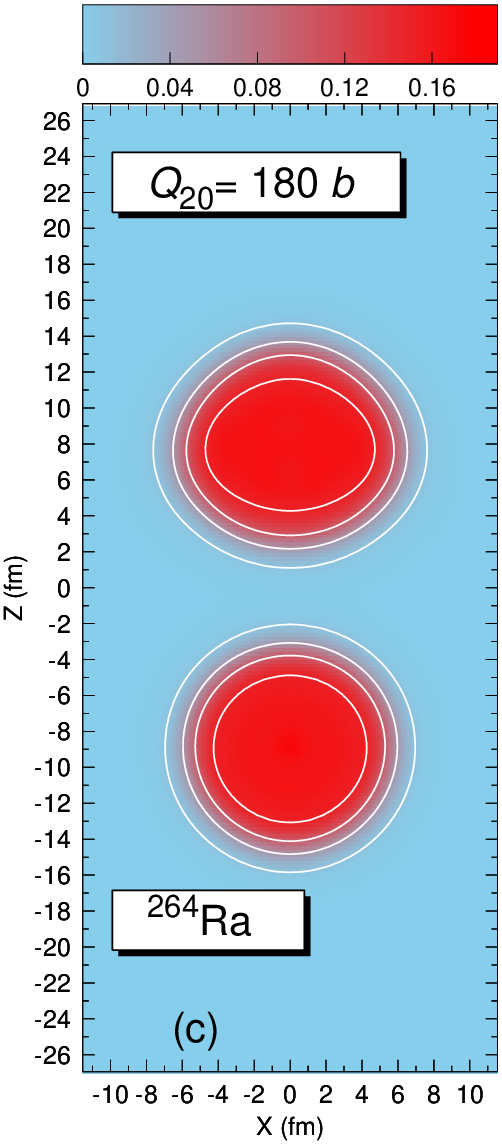}
\caption{ 
(Color online) Density contour plots for the nuclei
$^{236}$Ra [panel (a)],  $^{252}$Ra [panel b)] and  $^{264}$Ra [panel (c)].
The density profiles correspond to 2F solutions at the quadrupole 
deformations $Q_{20}$=150, 170 and 180 b, respectively. Results are 
shown for the parametrization D1M of the Gogny-EDF. The density is in
units of fm$^{-3}$ and contour lines
are drawn at 0.01, 0.05, 0.10 and 0.15 fm$^{-3}$. 
}
\label{den_cont_232-252-264Ra_D1M} 
\end{figure*}

In order to further understand the origin of the extra stability leading
to   second fission isomers, we have performed a similar analysis to 
the one of  \cite{Mcdonell-2} where the spatial density for that 
second isomer is compared to the densities of two nuclei summing up
the same number of protons and neutrons as the parent nucleus. In order
to choose the $N$ and $Z$ values used,  we plot the number of particles 
$N(z)=2\pi \int_{-\infty}^z dz' \int r_\perp dr_\perp \rho (r_\perp,z') $ 
up to a distance $z$. This quantity has several regions where it behaves
quadratically with $z$ and a central region (corresponding to the neck)
where it behaves almost linearly. By locating the mid point of that 
region of linear growing, we determine the $Z$ and $N$ values 
 mentioned above. A subsequent HFB calculation for the ground
state or an excited configuration (see below) allows to obtain the corresponding
densities that are then shifted as to match the tips of the density of the
parent nucleus. In Fig.~\ref{CDthird}, we have plotted the density 
(left hand side of each panel)
corresponding 
to the second fission isomer in $^{232}$Ra, $^{240}$Ra and $^{248}$Ra
and the densities (right hand side of each panel) of the 
two nuclei described above. In each panel the corresponding splitting
is given in the lower part. The  densities
computed with the D1S, D1N and D1M Gogny-EDFs are quite similar 
and therefore, we will focus on the results obtained with the D1S parameter 
set. The densities
of the three nuclei show some differences as a consequence of the different
number of protons and neutrons as well as the different quadrupole moment
but qualitatively they look rather similar pointing to a weak dependence
with neutron number. With respect to the "fragments", the heavier one
is always a rare earth nucleus with a prolate deformation in its ground
state. The light "fragment" is characterized by its proton number close or
equal to the magic $Z=28$ one, that leads to a spherical ground state. In spite
of being spherical, the three light "fragments" have a quadrupole 
potential energy surface
showing a shoulder at $\beta_2 \approx 0.25 - 0.35$ with an excitation energy of
just a couple  MeV. The density of that "excited configuration" matches
quite well the density of the second isomer and therefore the existence
of the third minimum could be linked to shell effects in the light system
associated to the above mentioned shoulder. This finding has to be confirmed
by a similar analysis in other Th, U and Pu nuclei showing a second fission isomer.

\begin{figure*}
\includegraphics[width=0.95\textwidth]{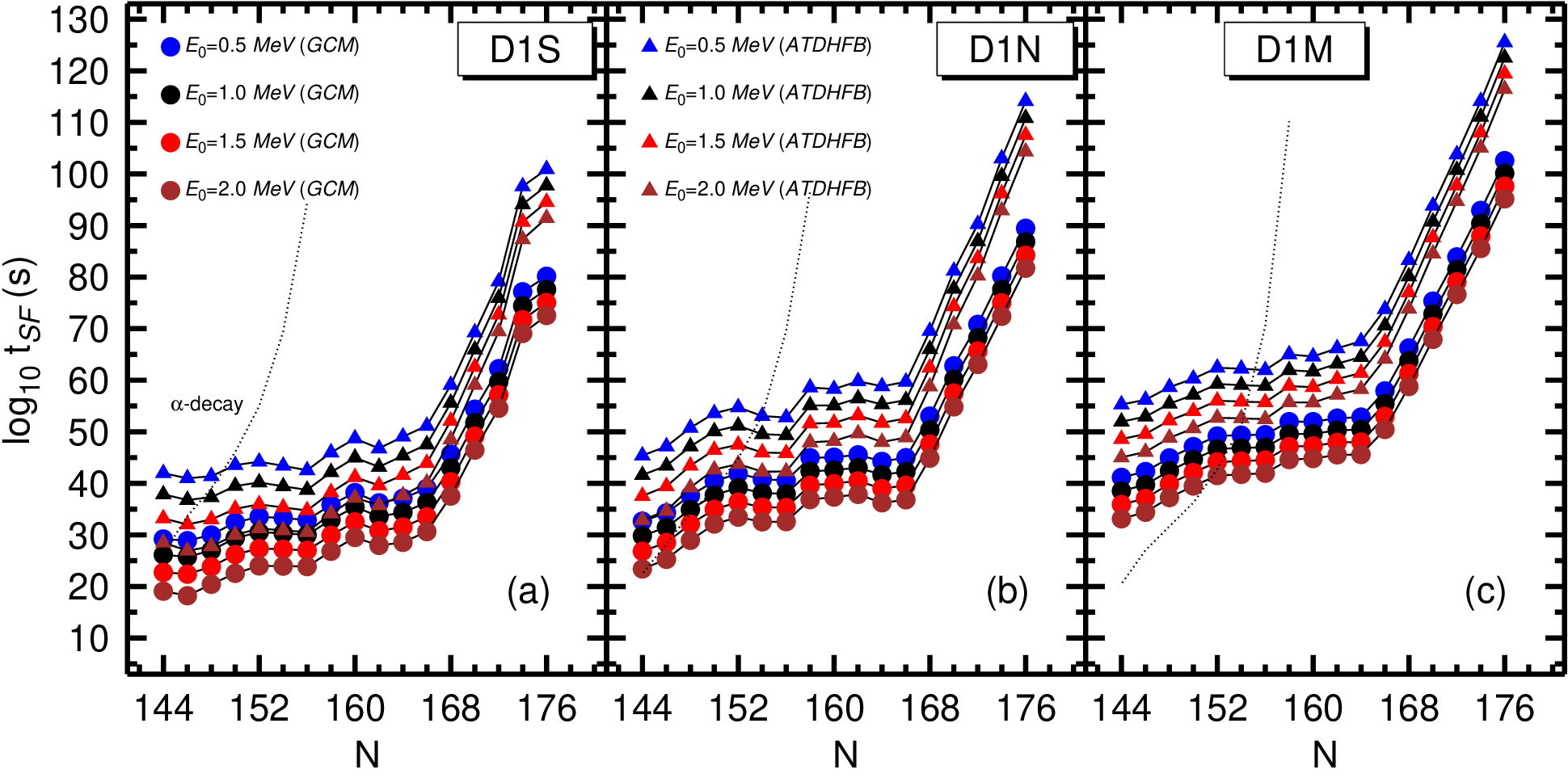}
\caption{ (Color online) The spontaneous fission half-lives t$_\mathrm{SF}$, 
predicted within the GCM and ATDHFB schemes, for the isotopes 
$^{232-264}$Ra  are depicted as functions of the neutron number. 
Results have been obtained with the Gogny-D1S [panel (a)], Gogny-D1N 
[panel (b)] and  Gogny-D1M [panel (c)] 
 EDFs. For each parametrization, 
calculations have been carried out  with E$_{0}$=0.5, 1.0, 1.5 and 
2.0 MeV, respectively. 
The  
$\alpha$-decay half-lives are also plotted with  short dashed lines. 
}
\label{tsf-D1S-D1N-D1M} 
\end{figure*} 

Let us turn our attention, to the proton ($Z_{1},Z_{2}$), neutron 
($N_{1},N_{2}$) and mass ($A_{1},A_{2}$) numbers of the fragments 
obtained at the HFB level in  $^{232-264}$Ra which are plotted in 
Fig.~\ref{mass-D1S-D1N-D1M}. As can be seen  the overall trend is quite 
reminiscent of the one in U  and Pu  nuclei 
\cite{Rayner-UPRC-2014,Rayner-UEPJA-2014}. The key role played by the 
magic neutron  and proton numbers in the fragments' masses and charges 
is also apparent from the figure. For example, exception made of the 
isotopes  $^{262,264}$Ra with the parametrizations D1S and D1N, the 
neutron number in one of the fragments is always rather close or equal 
to 82. Moreover,  for $^{232-244}$Ra, the proton number in one of the 
fragments is close to 50 which compares well with available 
experimental data for this region of the nuclear chart \cite{Schmidt}. 
We stress, however, that such a comparison with the experiment should 
be taken with care because, as already discussed above for the case of 
$^{240}$Ra, in our calculations the masses and charges of the  
fragments are obtained from HFB 2F solutions minimizing the energy at 
the largest quadrupole deformations.

In Fig.~\ref{den_cont_232-252-264Ra_D1M} we have plotted the 2F density 
contours for the nuclei $^{236}$Ra, $^{252}$Ra and  $^{264}$Ra at 
$Q_{20}$=150, 170 and 180 b, respectively. Results are shown for the 
Gogny-D1M EDF but similar ones are obtained with the D1S and D1N 
parametrizations. The lighter and heavier fragments in $^{236}$Ra and 
$^{264}$Ra  turn out to be oblate  ($\beta_{2}$=-0.23 and -0.14, 
respectively). In fact, our calculations predict, for example, oblate 
(-0.26 $\le$ $\beta_{2}$ $\le$ -0.17) and slightly octupole 
($\beta_{3}$ $\approx$ 0.03) deformed lighter fragments for the 
isotopes $^{232-244}$Ra. Such oblate fragments have also been found in 
previous HFB calculations based on both the Gogny and BCPM EDFs 
\cite{Rayner-UPRC-2014,Rayner-UEPJA-2014,Robledo-Giulliani}. A better 
understanding of the shell effects leading to them is required as only 
prolate deformations are usually assumed for the fission fragments 
\cite{Moller-1,Moller-2}. On the other hand, a symmetric splitting into 
two $^{126}$Ru nuclei  ($\beta_{2}$=-0.02, $\beta_{3}$ $\approx$ 0.01) 
is predicted for $^{252}$Ra. The same holds for the N=164 nuclei 
$^{256}$U \cite{Rayner-UPRC-2014} and $^{258}$Pu  
\cite{Rayner-UEPJA-2014} (two $^{128}$Pd and $^{129}$Ag fragments, 
respectively). Moreover, our calculations for $^{254}$Th suggest a 
splitting into two $^{127}$Rh fragments.

Finally, in Fig.~\ref{tsf-D1S-D1N-D1M}, we have depicted the 
spontaneous fission half-lives obtained for  $^{232-264}$Ra as 
functions of the neutron number. Both the GCM and ATDHFB schemes have 
been employed. For each isotope we have considered four values of  
E$_{0}$ (i.e., E$_{0}$=0.5, 1.0, 1.5 and 2.0 MeV) and, as can be seen 
from the figure, an increase of this parameter leads to a decrease in 
the t$_\mathrm{SF}$ values by several orders of magnitude. On the other hand, 
the ATDHFB half-lives are always larger than the GCM ones for a given 
E$_{0}$. For example, for  $^{234}$Ra ($^{252}$Ra) the GCM values are 
5.492 $\times$ 10$^{25}$ (2.243 $\times$ 10$^{34}$), 2.951 $\times$ 
10$^{31}$ (4.578 $\times$ 10$^{41}$) and 5.247 $\times$ 10$^{39}$ 
(2.755 $\times$ 10$^{50}$) s while ATDHFB ones are 5.884 $\times$ 
10$^{36}$ (2.293 $\times$ 10$^{45}$), 2.513 $\times$ 10$^{43}$ (1.992 
$\times$ 10$^{55}$) and  8.511 $\times$ 10$^{52}$ (3.094 $\times$ 
10$^{64}$) s with the D1S, D1N and D1M Gogny-EDFs and E$_{0}$=1.0 MeV. 
In fact, for the isotopes with N $\le$ 166, we have found differences 
of up to 11, 14 and 15 orders of magnitude between the two schemes. We 
observe a steady increase of the t$_\mathrm{SF}$ values in heavier isotopes, 
with the differences between the GCM and ATDHFB predictions reaching 
20, 24 and 22 orders of magnitude in the case of $^{264}$Ra with the 
D1S, D1N and D1M parameter sets (E$_{0}$=1.0 MeV). The conclusion is that
t$_\mathrm{SF}$ values strongly depend upon the details of the calculation and
the functional used with "error bars" of up to 20 orders of magnitude.
However, a global trend is observed in {\it all} the cases, i.e.,  the
slight increase of t$_\mathrm{SF}$ as a function of neutron number up to N=166 
which is followed by a  steeper increase that continues up to the largest
neutron number considered, ruling out 
the possibility of fission recycling in the r-process.
In order to examine the competition between the spontaneous fission and 
$\alpha$-decay modes, in Fig.~\ref{tsf-D1S-D1N-D1M} we have also 
included the $\alpha$-decay half-lives Eq.(\ref{VSeaborg-new}). Though 
the precise transition point depends of the selected Gogny-EDF, we 
conclude that with increasing neutron number fission turns out to be 
faster than $\alpha$-decay.

 As expected, pairing correlations have a 
strong impact on the spontaneous fission half-lives obtained for the 
considered Ra nuclei. We have tested that, similar to previous  Gogny 
and/or BCPM results 
\cite{Rayner-UPRC-2014,Rayner-UEPJA-2014,Robledo-Giulliani}, an 
increase of the pairing field by only 5 or  10 $\%$ leads to a 
significant decrease in the predicted t$_\mathrm{SF}$ values. However, even in 
such a case, the previous conclusion (i.e., fission dominates over 
$\alpha$-decay for increasing N) remains valid regardless of the 
Gogny-EDF used in the calculations.

\section{Conclusions}
\label{Coclusions}

In the present work, we have studied the fission properties of 
even-even Ra isotopes with neutron numbers  144 $\le$ N $\le$ 176 
within a mean field framework \cite{rs}. With the aim to test the 
robustness of our HFB predictions with respect to the particular 
version of the Gogny-EDF \cite{gogny} employed, calculations have been 
carried out with the parameter sets D1S \cite{gogny-d1s}, D1N 
\cite{gogny-d1n} and D1M \cite{gogny-d1m}. The fission paths (i.e., the 
1F and 2F solutions) have been determined, for each Ra isotope, with 
the help of constraints on the proton $\hat{Z}$ and neutron $\hat{N}$ 
number operators as well as on the axially symmetric quadruple 
$\hat{Q}_{20}$, octupole $\hat{Q}_{30}$, $\hat{Q}_{10}$ and necking 
$\hat{Q}_{Neck}(z_{0},C_{0})$ operators.  For each of the considered 
($Q_{20}$,$Q_{30}$,$Q_{Neck}$, $\dots$)-configurations we have also 
performed an optimization of the  lengths  of the (deformed) axially 
symmetric HO  single-particle basis in order to improve the convergence 
of the relative energies 
\cite{Rayner-UPRC-2014,Rayner-UEPJA-2014,Robledo-Giulliani}. The mean 
field equations have been solved using an approximate second order 
gradient method that allows to handle several constraints efficiently 
\cite{PRCQ2Q3-2012,Robledo-Rayner-JPG-2012,Robledo-Bertsch2OGM}. Zero 
point vibrational and rotational corrections have always been added to 
both the 1F and 2F  HFB energies.

Regardless of the considered Gogny-EDF, the 1F curves of the studied Ra 
nuclei exhibit a similar rich topography consisting of, for example, 
the ground state minimum, first and second fission isomers as well as 
first and second barriers. A change in tendency is observed for the 
excitation energies of the first $E_{I}$ and second  $E_{II}$  isomers 
as well as for the heights of the first $B_{I}$ an second $B_{II}$ 
barriers  at  the neutron number N=164.  For some selected Ra isotopes, 
we have also corroborated the expected reduction of the inner barrier 
heights $B_{I}$ \cite{Abusara-2010,Rayner-UPRC-2014,Rayner-UEPJA-2014} 
once the $\gamma$ degree of freedom is taken into account. Moreover, in 
order to better understand the structure of the third minima along the 
1F curves, we have studied their density profiles. Ours agree well with 
previous results \cite{Mcdonell-2} and suggest that those third minima 
in Ra nuclei could be linked  to shell effects associated with an 
excited (deformed) configuration in a light (spherical) system with Z 
$\approx$ 28. A more detailed study, including other Th, U and Pu 
nuclei showing a second fission isomer, will be presented elsewhere.

We have obtained the masses and charges of the fission fragments, from 
(variational) 2F solutions at the largest quadrupole moments. A 
symmetric splitting into two $^{126}$Ru nuclei is predicted for 
$^{252}$Ra as well as oblate and slightly octupole deformed fission 
fragments for several other Ra isotopes. Though the predicted overall 
trend for the fragments' masses and charges agrees reasonably well with 
available data \cite{Schmidt} for this region of the nuclear chart, we 
stress that our procedure tends to overestimate 
\cite{Rayner-UPRC-2014,Rayner-UEPJA-2014} the role of the proton Z=50 
and neutron  N=82 magic numbers 
\cite{Nenoff-2007,Piessens-1993,Ter-1996}. Therefore, a more 
sophisticated \cite{Goutte-dynamical-distribution} approach, taking 
into account the quantum dynamics around the scission configurations 
\cite{Chasman-breaking}, is still required.

The spontaneous fission half-lives t$_{SF}$ have been computed within 
the standard WKB approximation \cite{Baran-TSF-1,Baran-TSF-2}. Both the 
ATDHFB \cite{crankingAPPROX,Giannoni-1,Giannoni-2,Libert-1999} and the 
GCM \cite{rs,Rayner-UPRC-2014,Rayner-UEPJA-2014,Robledo-Giulliani} 
schemes have been employed to obtain the collective masses and the zero 
point vibrational corrections while for the rotational energies we have 
resorted to an approximate variation-after-projection (VAP) \cite{rs} 
in terms of the Yoccoz moment of inertia \cite{RRG23S,ER-Lectures}. In 
spite of the large uncertainties in the predicted t$_\mathrm{SF}$ values, 
mainly related to the details of the calculations (including the 
strength of pairing correlations and the Gogny-EDF  used) a robust 
global trend is observed, i.e.,  the slight increase of the t$_\mathrm{SF}$ 
values  up to N=166 followed by a  steady increase up to  $^{264}$Ra 
that correlates well  with the one of the barrier heights and the 
widening of the 1F curves in heavier Ra isotopes. As a result, heavier 
neutron-rich Ra nuclei become stable against spontaneous fission, 
diminishing the importance of fission recycling in the r-process 
\cite{Pan05,Mar07,Pan08}. Furthermore, our calculations indicate that 
with increasing neutron number fission turns out to be faster than 
$\alpha$-decay.

From the results discussed in this paper we conclude that the fission 
properties found for neutron-rich U and Pu  
\cite{Rayner-UPRC-2014,Rayner-UEPJA-2014,Robledo-Giulliani} are  
preserved down to neutron-rich  Ra nuclei. This motivates further 
explorations in this region of the nuclear chart using the Gogny-HFB 
framework. Within this context, a long list of tasks should be 
undertaken. Among them, the following two appear as our next plausible 
steps. First, a minimal action, instead of minimal energy, description 
of  fission including pairing fluctuations \cite{Action-Rayner} in 
addition to multipole moments \cite{deltaS-Doba} should receive more 
attention. However, a more realistic treatment of the vibrational mass 
parameters is still required within the framework presented in 
\cite{Action-Rayner}. Here, we also refer the reader to the recent 
study \cite{RPA-masses} where  coherent and time-feasible calculations 
of the vibrational masses have been envisioned using the Gogny-HFB plus 
the quasiparticle random-phase approximation (RPA). Second, the study 
of the fission properties in odd-A nuclei will provide valuable 
information on the predictive power of our Gogny-HFB approach to 
account for the larger t$_\mathrm{SF}$ values in those nuclear systems 
compared with the corresponding even-A neighbors (i.e., the hindrance 
factors) \cite{Bjor}. Here, the Gogny-HFB equal filling approximation 
(EFA) \cite{perez,ours_plb,EFA-o-1}, represents a reasonable and 
computationally feasible starting point. Work along these lines is in 
progress.

\begin{acknowledgments}
The work of L. M. Robledo has been supported in part by MINECO grants Nos. FPA2012-34694, 
FIS2012-34479 and by the Consolider-Ingenio 2010 program MULTIDARK 
CSD2009-00064. 
\end{acknowledgments}

\end{document}